# TOWARDS ALGORITHM-FREE PHYSICAL EQUILIBRIUM MODEL OF COMPUTING

S. Mousavi
COVENTRY UNIVERSITY
*ad0204@coventry.ac.uk*

Our computers today, from sophisticated servers to small smartphones, operate based on the same computing model, which requires running a *sequence* of discrete instructions, specified as an algorithm. This sequential computing paradigm has not yet led to a fast algorithm for an NP-complete problem despite numerous attempts over the past half a century. Unfortunately, even after the introduction of quantum mechanics to the world of computing, we still followed a similar sequential paradigm, which has not yet helped us obtain such an algorithm either. Here a completely different model of computing is proposed to replace the sequential paradigm of algorithms with inherent parallelism of physical processes. Using the proposed model, instead of writing algorithms to solve NP-complete problems, we construct physical systems whose equilibrium states correspond to the desired solutions and let them evolve to search for the solutions. The main requirements of the model are identified and quantum circuits are proposed for its potential implementation.

*Keywords*: NP-completeness, Quantum computing, Equilibrium states

## 1  Introduction

Since the first NP-completeness proof in 1971[1], countless problems have been identified as NP-complete, but no fast algorithm has yet been found for them[2]. By a fast (or quick) algorithm, we mean a polynomial-time algorithm, i.e. an algorithm whose running time grows polynomially with input size. Unfortunately, even the stunning properties of quantum mechanics have not yet resulted in such algorithms for NP-complete problems. The computing model proposed in this paper is algorithm-free and relies on physical processes to solve decision problems (which include NP-complete problems). A decision problem only has two possible answers of positive (Yes or True) or negative (No or False).

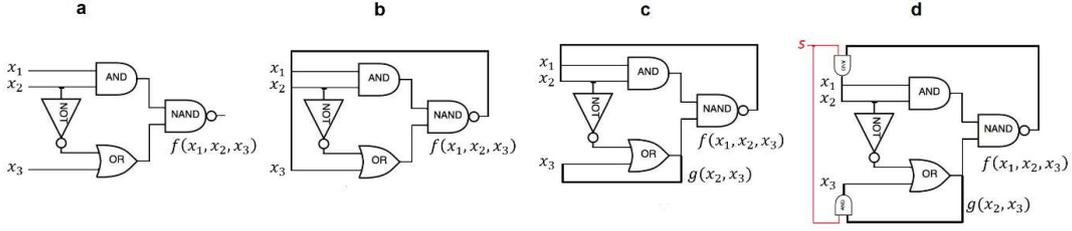

**Fig.1 | Circuits of logic gates.**
**a**, A circuit with three external inputs for computing the function $f(x_1, x_2, x_3) = \text{NAND}(\text{AND}(x_1, x_2), \text{OR}(\text{NOT}(x_2), x_3))$. **b**, The circuit obtained by feeding back the output of the NAND gate in the circuit shown in part **a** to all its external inputs. This circuit is closed because it has feedback loop and no external input. **c**, Another closed circuit obtained from the circuit shown in part **a** which is the same as the circuit in part **b** except that it uses a different feedback for $x_3$. **d**, This circuit is not closed because it has an external input s. However, it will be equivalent to the closed circuit shown in part **c** when its input s is set to 1, because s will be fixed as if it is no longer an external input and the small AND gates in the figure will act as wires. Therefore, the input s may be used to control when the circuit behaves as the closed circuit of part **c**. Such a technique may be applied to other closed circuits, by adding a control input and a number of AND gates, to start and stop their closeness behaviour.

The main idea behind the proposed model is to construct a physical system $s$ for a given instance $I$ of a decision problem such that the answer to $I$ is positive iff $s$ has an equilibrium (stable) state. Therefore, the proposed model is called the Physical Equilibrium Computing (PEC) model. The model has requirements to meet, which are identified in this paper, and needs physical implementation to actually solve NP-complete problems.

The rest of this paper is organised as follows. In Section 2, three new concepts of closed circuits, Closed-Circuit Stability (CCS) problem, and physical closed circuits are introduced. Based on these concepts, Section 3 presents the PEC model. Section 4 compares the proposed model with similar models in the literature, and Sections 5 concludes the paper.

## 2  Basic definitions

A circuit of a universal set of logic gates can represent any Boolean function, from $\{0,1\}^n$ to $\{0,1\}^m$, where $n \geq 0$ is the number of external inputs to the circuit and $m \geq 1$ is the number of its outputs, which is considered 1 in the rest of the paper (Fig. 1.a).

**Definition 1.** By a *closed circuit*, we mean a circuit with feedback loop(s) and no external input (Fig.1.b and 1.c.).

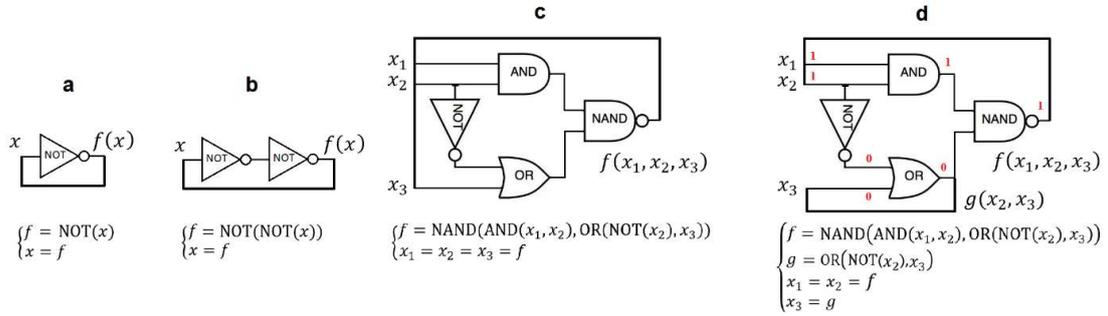

**Fig. 2 | Examples of stable and unstable closed circuits and their *characteristic systems***
**a**, This is the simplest unstable circuit. As can be seen, the output of the NOT gate is connected to its input. Therefore, no Boolean value (0 or 1) can be assigned to its output so that its inversion functionality is preserved. The behaviour of the circuit is described by its characteristic system, displayed underneath the circuit, which is a system of Boolean equations to specify both feedforward and feedback relations. This characteristic system is infeasible, i.e. has no solution in the Boolean domain, which confirms the instability of the circuit. **b**, This circuit is stable because its characteristic system is feasible. In fact, both $x = 0$ and $x = 1$ are solutions to the characteristic system. **c**, This circuit, also shown in Fig.1.b, is unstable because its characteristic system is infeasible. This means there is no assignment of Boolean values to the outputs of the gates such that every gate functions as expected. **d**, This circuit, previously seen in Fig. 1.c, is stable because its characteristic system is feasible. Indeed, its solution is unique, $x_1 = x_2 = 1, x_3 = 0$, for which the gates' outputs are also displayed. As can be seen, the functionality of every gate is preserved by these values.

Given a potential assignment of Boolean values to the gates' outputs in a closed circuit, the corresponding gates' input values will be uniquely identified (via the feedback loops) because there is no external input. However, the potential input and output values assigned to a gate do not necessarily match its actual function, in which case its functionality is not preserved by these values. A wire may also be viewed as a gate with the identity function.

**Definition 2.** Given a closed circuit $c$, the *Closed-Circuit Stability* problem is the problem of deciding whether there exists an assignment of Boolean values to the outputs of the gates in $c$ such that the Boolean functionality of every gate is preserved. The closed circuit $c$ is called *stable* if such an assignment exists and *unstable* otherwise (Fig. 2).

**Definition 3.** By a *physical closed circuit*, we mean a physical system that implements a closed circuit. It is called *stable* if its gates' output values remain fixed (under normal conditions) and preserve the functionality of every gate; it is called unstable otherwise.

As physical systems, the gates and wires in a physical circuit may be subject to malfunctioning or delayed functioning.

In the rest of the paper, for simplicity and to distinguish between closed circuits and their physical implementation, we may refer to them as *logical circuits* and *physical circuits*, respectively. For brevity, by the *output values* of a (logical or physical) circuit, we mean the output values of its gates.

## 3  The proposed model

In this section, first, the requirements for solving the CCS problem using physical circuits are identified and further requirements for this solution to be quick are explored. Then, this quick solution of CCS is applied to quickly solve all NP-complete problems. Next, a method based on the concept of semi-closed circuits is introduced to reuse the same circuit to solve problem instances of the same size, and some related properties are explored. Then, a method is outlined to verify the stability of a physical circuit by observing a subset of its gates. Finally, quantum circuits are proposed for further investigation to implement physical circuits.

### 3.1. Quick solution of CCS

The central idea behind the PEC model is to construct physical circuits to solve the CCS problem. More specifically, to determine if a given logical circuit $c$ is stable, we implement it as a physical circuit $s$ and see if $s$ becomes stable. We will solve other decision problems by recasting them as CCS. However, to solve CCS using physical circuits we need to make the following assumptions.

**Assumption 1** (*Construction*). We can construct a physical circuit $s$ to implement a given logical circuit $c$.

**Assumption 2** (*Stability*). Given Assumption 1, the physical circuit $s$ will become stable within some definite time limit iff $c$ is stable.

**Assumption 3** (*Verification*). Given Assumption 2, we can observe the output values of $s$ (to verify the stability of $c$).

These assumptions are now elaborated on. Assumption 1 is valid if the following requirements are met:

- *Requirement* 1.i. We can physically construct NAND (or any set of universal) gates with only two possible *states* as their input and output Boolean values; no third state must be possible. For example, conventional electronic gates cannot be used because they allow (voltage) values outside the valid ranges for Boolean 0 and 1. These two possible states are also denoted by 0 and 1, for simplicity.

- *Requirement* 1.ii. We can 'wire' the output $y$ of any gate to one or more inputs $x$ of the same or other gates so that the state of $x$ will be equal to that of $y$.

By Definition 3, the output values of a stable physical circuit preserve the functionality of every gate in $s$ and, equivalently, every gate in its logical circuit. Therefore, the stability of a physical circuit implies the stability of its logical circuit (by Definition 2), i.e. the 'only if' direction of Assumption 2 is valid. Its 'if' direction is also valid if each physical circuit $s$ that implements a stable logical circuit $c$ becomes stable within a definite time limit specified, for example, as a function of its size $|s|$.

Assumption 3 only requires our ability to observe the output values of a physical circuit $s$. Once we have these values (observed after the definite time mentioned in Assumption 2), we can use them to verify the stability of the corresponding logical circuit $c$. More specifically, these values preserve the functionality of every gate in $c$ iff $c$ is stable (by Definitions 2-3 and Assumption 2). A method is outlined in Section 3.4 to reduce the number of gates we need to observe for this purpose.

We are now ready to present the first result. We say that a decision problem $p$ can be solved by the PEC model, or is *PEC-decidable,* if the answers to its instances can be obtained by constructing physical circuits and observing their outputs (after some definite time). The problem is *quickly* PEC-decidable if it can be solved by the PEC model quickly, i.e. in polynomial time.

Given an instance $c$ of CCS, we can construct the corresponding physical circuit $s$, by Assumption 1, which will become stable by some definite time iff $c$ is stable, by Assumption 2. We can observe the output values of $s$ to verify the stability of $c$, by Assumption 3. This gives the following result.

**Proposition 1.** Under Assumptions 1-3, CCS is PEC-decidable.

Given Proposition 1, it is not hard to see that every decision problem solvable by a Turing machine (or, equivalently, by current conventional computers) is also PEC-decidable under Assumptions 1-3 (shown in Section 3.3). However, the main motivation for proposing the PEC model is not to solve such problems but to investigate its potential to quickly solve NP-complete problems. Such a quick solution cannot be concluded from Assumptions 1-3 alone, because these assumptions are concerned with our *ability* to verify the stability of logical circuits (via the construction of their physical circuits) irrespective of the *time* needed to do so.

**Assumption 4** (*Polynomial time boundedness*). The construction, stability, and stability verification of a physical circuit corresponding to a logical circuit $c$ are performed in polynomial time in $|c|$.

More specifically, Assumption 4 requires the existence of a polynomial $p$ such that for each logical circuit $c$, the time needed to construct the corresponding physical circuit $s$, the time needed for $s$ to become stable if $c$ is so, and the time needed to observe its output values are less than $p(|c|)$. This assumption implies that $|s|$ is also polynomial in $|c|$. Using Assumption 4, Proposition 1 is extended to the following result.

**Proposition 2.** Under Assumptions 1-4, CCS is quickly PEC-decidable.

*3.2. Quick solution of NP-complete problems*

To extend the quick solution of CCS (Proposition 2) to other NP problems, we now prove that CCS is NP-complete.

**Theorem 1.** CCS is NP-complete.

**Proof.** It is not hard to see that CCS is NP by using as certificate $y$ an assignment of Boolean values to the outputs of the gates in a given instance $c$. The size of the certificate is $O(|c|)$, and it takes polynomial time to check if such an assignment $y$ preserves the functionality of every gate in $c$.

We will now show that the Circuit Satisfiability (CSAT) problem is quickly reducible to CCS. Let $I$ be an instance of CSAT, i.e. a circuit of Boolean gates with $n > 0$ external inputs $x_1, \ldots, x_n$ that computes a Boolean function $f(x_1, \ldots, x_n)$. We can construct in polynomial time (in $|I|$) a closed circuit $c$ by adding to $I$ one NOT and $n$ XOR gates $g_i, i = 1, \ldots, n$, and making the following connections. The input of the extra NOT gate is connected to the output of the circuit $I$. For each extra XOR gate $g_i, i = 1, \ldots, n$, an input of $g_i$ is connected to the output of the extra NOT gate, and its other input and its output are both connected to the external input $x_i$ (see Fig. 3).

We now show that the resulting closed circuit $c$ is stable iff the instance $I$ of CSAT is satisfiable. First assume that $c$ is stable. This means there exists an assignment of Boolean values to the outputs of the gates in $c$ such that the functionality of every gate is preserved. For each $i = 1, \ldots, n$, let $v_i$ be the value assigned to the output of the XOR gate $g_i$. Because the output value of $g_i$ is the same as one of its input values, its other input value must be 0; otherwise, its exclusive-or functionality would have not been preserved. This implies $f(v_1, \ldots, v_n) = 1$, i.e. $I$ is satisfiable. Now assume that $I$ is satisfiable. Then, there exists an assignment $x_i = v_i, i = 1, \ldots, n$, that satisfies $I$. Obviously, for this assignment, while the functionality of all its gates is preserved, the circuit $I$ outputs $f(v_1, \ldots, v_n) = 1$. Therefore, $v_i = \text{XOR}\left(v_i, \text{NOT}(f(v_1, \ldots, v_n))\right), i = 1, \ldots, n$. That is, the functionality of the extra NOT and XOR gates is also preserved by this assignment, so $c$ is stable. ∎

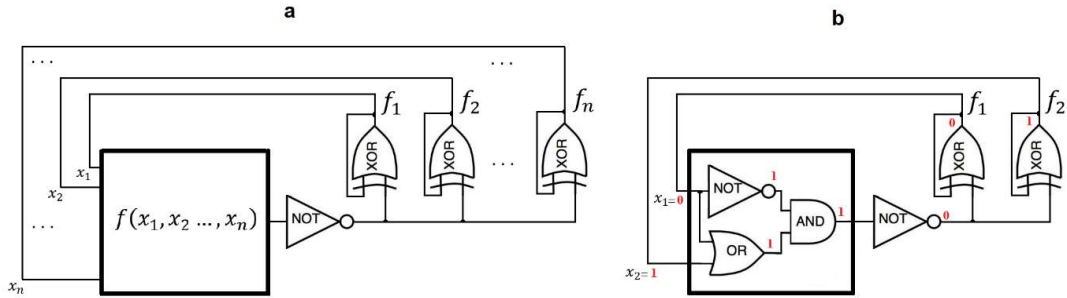

**Fig. 3 | How to construct an instance of CCS, i.e. a closed circuit, from a given instance of CSAT.**
**a**, The closed circuit is obtained by adding one NOT and $n$ XOR gates to the given instance $I$ of CSAT shown as a black box (with thick border) that computes a Boolean function $f(x_1, \ldots, x_n)$. As can be seen, the external inputs $x_i, i = 1, \ldots, n$, to the black box are now provided by the feedback loops from the outputs of the added XOR gates, so the resulting circuit $c$ is closed. **b**, As an example, the resulting closed circuit $c$ is illustrated for a given circuit $I$ that computes $f(x_1, x_2) = \text{AND}(\text{NOT}(x_1), \text{OR}(x_1, x_2))$. The given circuit $I$ is satisfiable because there exists an assignment to its external inputs ($x_1 = 0, x_2 = 1$) for which it outputs True (1). As can be seen this satisfying assignment is consistent with the functionality of every gate in $c$ whose output values are shown, hence $c$ is stable. Indeed, the assignment ($x_1 = 0, x_2 = 1$) is the only satisfying assignment for $I$. That is, there are no other potential output values consistent with the functionality of all the gates in $c$.

Theorem 1 implies, given an instance $I$ of an NP-complete problem, we can quickly formulate $I$ as an equivalent instance $c$ of CCS. This result, together with Proposition 2, yields the following result.

**Proposition 3.** Under Assumptions 1-4, every NP-complete problem is quickly PEC-decidable.

*3.3. Reuse of physical systems*

This section is not a core part of the proposed model and may be skipped.

Theorem 1 implies that an instance $I$ of an arbitrary NP-complete problem $x$ may be solved by formulating $I$ as an instance $c$ of CCS and solving $c$ via the construction of a physical system $s$. However, the construction of $s$ from $I$ is *per-instance*, i.e. $s$ varies with $I$. This per-instance construction of physical systems is prohibitive in practice (as is the case with other natural computing proposals). To mitigate this issue in PEC, it is shown, at least in theory, that we can construct systems *per-instance-size*. More specifically, we can reuse the same system to solve all the instances of the same size.

**Definition 4.** A *semi-closed circuit* is a circuit of Boolean gates with both feedback loop(s) and external input(s).

Elimination of all external inputs of a semi-closed circuit by fixing them to arbitrary Boolean values will result in a closed circuit (by Definition 1).

**Definition 5.** Let $i_1 \ldots i_n$ denote the bit string representing an instance $I$ of a decision problem and $c$ be a semi-closed circuit with $n = |I|$ external inputs $x_1, \ldots, x_n$. We say $I$ is *solvable* by $c$ provided that the answer to $I$ is positive iff $c_I$ is stable, where $c_I$ is the closed circuit obtained by fixing each external input $x_k$ to $i_k$, $k = 1, \ldots, n$.

**Theorem 2.** Every problem $x$ in NP can be solved by a family $F$ of polynomial-size semi-closed circuits. That is, there exists a polynomial $p$ such that for each $n \geq 0$, there exists a semi-closed circuit of size no more than $p(n)$ that solves every instance $I$ of $x$ such that $|I| = n$.

**Proof.** Each NP problem $x$ has a polynomial-time verifier $v$ and a polynomial $q$ that bounds the certificate size. Let $I$ be an instance of $x$. The verifier $v$ accepts a given pair $(I, y)$ iff $y$ is a valid certificate for $I$ (in which case $|y| \leq q(|I|)$). Assume wlog the input alphabet is $\Sigma = \{0,1\}$ and $q(|I|) > 0$. Then, there exists a polynomial-size circuit $c$ with $n$ external inputs $i_1, \ldots, i_n$ for the instance $I$, where $n = |I|$, and $s$ external inputs $y_1, \ldots, y_s$ for a candidate certificate $y$ of maximum size $s = q(|I|)$ that outputs 1 iff $y$ is a valid certificate for $I$. Let $f$ be the function that $c$ computes. Using one NOT and $s$ XOR gates, we can add to $c$ the feedback loops $y_k = \text{XOR}(y_k, \text{NOT}(f(i_1, \ldots, i_n, y_1, \ldots, y_s)))$, $k = 1, \ldots, s$, to obtain a semi-closed circuit $c_n$ (with $n$ external inputs and $s$ feedback loops) that solves $I$. Because no assumption has been made on the instance $I$ of $x$ except that its size is $n$, the semi-closed circuit $c_n$ solves every instance $J$ of $x$ such that $|J| = n$. The required family is $F = \{c_n, n \geq 0\}$. ∎

**Corollary.** Every decision problem $x$ with a verifier and a size-bounded certificate is solvable by a family of semi-closed circuits.

This corollary is obtained by using the same proof as used for Theorem 2 except that the verifier's running time, the certificate size, and the resulting circuit size are no longer required to be polynomially-bounded. It implies that every decision problem $x$ solvable by a Turing machine $M$ is also PEC-decidable because $M$ can be used (trivially) to construct a verifier for $x$ with a dummy (nonempty) string as the certificate.

In analogy to the P/poly complexity class, let SCP/poly be the complexity class of decision problems solvable by a family of polynomial-size semi-closed circuits. Then NP ⊆ SCP/poly by Theorem 2. This result is especially interesting because it means NP-complete problems have polynomial-size semi-closed circuits while they do not have polynomial-size circuits under the well-known conjecture NP ⊄ P/poly. We do not know at the time of writing whether the converse of Theorem 2 also holds (i.e.

SCP/poly = NP) but we conjecture not. We do not know if BQP ⊆ SCP/poly either, but we conjecture so and leave it as a future research.

*3.4. Stability verification with a reduced number of observed gates*

In this section, to facilitate Assumption 3, a method is outlined for stability verification of a logical circuit $c$ and, equivalently, its physical circuit $s$ by observing a subset of the gates in $s$.

The characteristic system of $c$ can be written as a set of $m > 0$ feedforward equations $y_j = f_j(x_1, \ldots, x_n)$, $j = 1, \ldots, m$, where $f_j$ is a Boolean function of (at most) $n > 0$ variables, and $n$ feedback equations $x_i = y_{k_i}$, $1 \leq k_i \leq m$, $i = 1, \ldots, n$ (see Fig. 2 for examples of feedforward and feedback equations). If $c$ is stable, $s$ will be stable by some definite time $t$. Let $v_{k_i}$ be the output value (observed after time $t$) that corresponds to $y_{k_i}$, $i = 1, \ldots, n$. It is not hard to see that these $n$ output values are sufficient to verify the stability of $c$ (and $s$). To that end, we show that $c$ is stable iff the equations $v_{k_i} = f_{k_i}(v_{k_1}, \ldots, v_{k_n})$, $i = 1, \ldots, n$, hold. First assume that $c$ is stable. Because of the stability of $s$ at the observation time, the observed values $v_{k_i}$, $i = 1, \ldots, n$, must satisfy the characteristic system of $c$, which implies $v_{k_i} = f_{k_i}(v_{k_1}, \ldots, v_{k_n})$, $i = 1, \ldots, n$. Now assume that these equations hold. Then the assignment $x_i = v_{k_i}$, $i = 1, \ldots, n$ will be a valid solution to the characteristic system, i.e. $c$ is stable. Therefore, it is enough to observe the outputs of $n$ gates, where $n$ is the number of feedback equations, and compute the same number of feedforward functions to verify the stability of $s$.

*3.5. Implementation of PEC in quantum systems*

The PEC model is abstract and requires physical implementation. That is, to quickly solve NP-complete problems, we need to build physical circuits so that Assumptions 1-4 are met. At the time of writing, we do not know how to achieve such implementation. However, because of its potential significance, it is worth investigation. One possibility, subject to further research, is to employ quantum circuits because of existing (and foreseeable) technologies for their construction and their potential to meet the requirements. Using quantum circuits in the unitary model has not yet resulted in quick algorithms for NP-complete problems. However, their usage in the PEC model will be fundamentally different, if the requirements are met.

Because of the existing technology to build universal quantum gates, e.g. Toffoli and Fredkin, Requirement 1.i is met. Whether Requirement 1.ii is also met needs further investigation, especially for unstable circuits. However, quantum wires are built and research is progressing on (coherent) quantum feedback[3-5]. It is also possible to observe the outputs of the gates in quantum circuits, so Assumption 3 holds. The validity of Assumptions 2 and 4 is subject to investigation. However, informally speaking, it is reasonable to expect the validity of Assumption 2 and the first and the last requirements of Assumption 4, i.e. the polynomial-time construction and stability verification of physical circuits. That would leave the polynomial-time stability of physical systems (when the corresponding logical circuits are stable) as a key requirement for this model to quickly solve NP-complete problems. Whether this requirement is met would need further investigation in particular real experiments because the interrelated physical processes that enforce the evolution of a physical system towards its equilibrium may not be fully captured via simulation.

## 4 Comparison with similar models

If implemented in quantum systems, the PEC model will be similar to the adiabatic quantum computing (AQC) model and its variants[6,7]. A common feature of these models is to utilise the parallel quantum dynamics as opposed to the sequential application of quantum gates in the unitary model. However, they are different in several aspects.

The fundamental distinction is that the sequential algorithmic paradigm, which is eliminated in PEC, is still an inherent element of AQC. More specifically, in AQC, *we* are responsible for specifying a sequence of steps (the algorithm) to gradually move the state of the system towards a solution state. Devising such algorithms has been a research question since AQC was proposed. Even under ideal physical conditions, how slowly to run the algorithm is unknown (as it depends on the energy gap). Indeed, AQC (with non-stoquastic Hamiltonian) is shown to be polynomially-equivalent to the unitary model[8], which implies it cannot quickly solve NP-complete problems unless so can the unitary model. In contrast, the PEC model, once implemented so that its requirements are met, is algorithm-free. That is, the system itself is responsible for its evolution, enforced by interplaying physical processes (without any need for us to specify them). This shift of responsibility from us to the system replaces the sequential algorithmic paradigm with inherently parallel processes. In addition, relying on natural processes makes PEC tolerant to the natural level of noise, whereas AQC follows a certain model (described by Schrodinger's equation) under noise-free conditions.

Another, less important, difference is in the type of computational problems they solve, which is optimisation for AQC and decision for PEC. AQC may yield a suboptimal solution. To check the optimality of its solutions for an NP-hard optimisation problem is an NP-complete problem on its own. It takes polynomial time to check the validity of solutions provided by PEC for an NP-complete problem.

Another similar work is the balanced machine model[9]. It also aims to exploit the stable (balanced) states of physical machines to perform computation. However, it uses physical systems with continuous states to implement logic functions. To the best of our knowledge, the implementation of logic functions requires physical systems with inherent discrete states, as emphasised in Requirement 1.i.

## 5 Conclusion

In this paper, an abstract computing model, fundamentally different from our conventional and current quantum computing models, was proposed to quickly solve NP-complete problems, its requirements were identified, and quantum circuits were proposed for its potential implementation. The model was compared with some similar works in the literature.

Due to the high potential of (current and foreseeable) quantum systems to meet the requirements and the potential of the proposed model to quickly solve NP-complete problems for the first time, the model and its implementation are worth further investigation. It is important to emphasise that real experiments are needed to investigate the time a physical circuit takes to reach its potential equilibrium, and not all the interrelated physical processes may be captured by simulation.

**Acknowledgement**

The author would like to thank Animesh Datta, Seyed S. Mousavi and Damien Foster for their comments and Nik Tsanov for his help with implementing circuits in FPGAs.